# Theoretical Expectations for Bulk Flows in Large Scale Surveys

Hume A. Feldman and Richard Watkins

*Physics Department, University of Michigan*
*Ann Arbor, MI 48109*

## ABSTRACT

We calculate the theoretical expectation for the bulk motion of a large scale survey of the type recently carried out by Lauer and Postman. Included are the effects of survey geometry, errors in the distance measurements, clustering properties of the sample, and different assumed power spectra. We consider the power spectrum calculated from the IRAS–QDOT survey, as well as spectra from hot + cold and standard cold dark matter models. We find that sparse sampling and clustering can lead to an unexpectedly large bulk flow, even in a very deep survey. Our results suggest that the expected bulk motion is inconsistent with that reported by Lauer and Postman at the $90 - 94\%$ confidence level.

*Subject headings*: cosmology: theory – observation – galaxies: clustering – distances and redshifts

## 1. Introduction

Recently, Lauer and Postman (1993) (LP) measured the bulk motion of a volume limited complete sample of all 119 Abell clusters with redshifts out to $15,000$ km/s. They used a distance indicator based on brightest cluster galaxies that gives a typical distance error of 16% for a single cluster. They report that the frame defined by their sample exhibited a bulk motion of $689 \pm 178$ km/s with respect to the cosmic rest frame defined by the CMB. This value seems quite large given current expectations for the power spectrum on these scales. However, as we shall see below, the Abell clusters are a sparse sampling of the underlying peculiar velocity field, in the sense that the intracluster spacing is large compared to the scale on which the field has significant variations. This implies that the bulk motion of the Abell cluster sample is not necessarily indicative of the motion of the volume as a whole.

In this letter, we study the theoretical expectations for a measurement of the type made by Lauer and Postman. Given a catalog of cluster positions and an assumed power spectrum, we calculate the covariance matrix for a measurement of bulk motion in the sample using a linear analysis based on the work of Kaiser (1989). We show that the covariance matrix is a sum of a noise term and the convolution of the velocity power spectrum with a window function. Given that the covariance matrix is roughly diagonal and isotropic, we can calculate the expectation value for the bulk flow as well as the probability for measuring a velocity greater than some reference value. We consider power spectra from the IRAS–QDOT survey (Feldman, Kaiser & Peacock 1993), the BBKS CDM model (Bardeen *et al.* 1986) and from simulations of hot + cold dark matter (HCDM) (Klypin *et al.* 1993). For our analysis we use the Lauer and Postman sample as well as simulated cluster catalogs designed to explore the dependence of our results on the radius and clustering properties of the survey. It should be noted that we treat the power spectrum and cluster catalogs as being independent. This allows us to avoid questions of whether the clustering properties of Abell clusters are representative of clusters as a whole (Efstathiou *et al.* 1992). In fact, we show that the high degree of clustering in the Abell cluster sample results in a significant increase of the expected bulk motion. Preliminary results of this analysis were presented elsewhere (Feldman & Watkins 1993). Our analysis can be contrasted with that of Strauss *et al.* (1993) who analyzed Monte-Carlo realizations of peculiar velocity data drawn from N-body simulations of specific theoretical models.

## 2. Analysis

The maximum likelihood solution for the uniform streaming motion $U_i$ of a cluster sample (Dressler *et al.* 1987) is given by

$$U_i = A_{ij}^{-1} \sum_n \frac{r_{n,j} S_n}{\sigma_n^2 + \sigma_*^2} , \qquad (1)$$

where $r_n$ is the position of the $n$th cluster, $S_n$ is its estimated line–of–sight velocity, $\sigma_*$ is the 1$D$ velocity dispersion, $\sigma_n$ is the estimated uncertainty in the line of sight peculiar velocity, and

$$A_{ij} = \sum_n \frac{\hat{r}_{n,i}\, \hat{r}_{n,j}}{\sigma_n^2 + \sigma_*^2} . \qquad (2)$$

$\sigma_n$ is proportional to the distance and is denoted by $\sigma_n = (\beta/100)r_n$, where $\beta$ is the percentage error in the distance measure. For the LP survey $\beta = 16\%$. In these equations and the equations to follow, repeated indices denote implicit sums. The estimated line-of-sight peculiar velocity $S_n$ is related to the true velocity by

$$S_n = \hat{r}_{n,i}\, v_i(\vec{r}_n) + \varepsilon_n , \qquad (3)$$

where $\varepsilon_n$ is drawn from a Guassian with zero mean and variance $\sigma_\varepsilon^2 = \sigma_n^2 + \sigma_*^2$. For the catalogs of interest $A_{ij}$ is nearly diagonal, with the off–diagonal terms being of order 10% of the diagonal ones.

Following Kaiser (1989) we write the uniform bulk velocity in terms of Eqs. (1), (2) and (3):

$$U_i = U_i^{(v)} + U_i^{(\varepsilon)} = A_{ij}^{-1} \sum_n \frac{\hat{r}_{j,n} \hat{r}_{k,n}}{\sigma_n^2 + \sigma_*^2} v_k(\hat{r}_n) + A_{ij}^{-1} \sum_n \frac{\hat{r}_{j,n} \varepsilon_n}{\sigma_n^2 + \sigma_*^2} \ . \tag{4}$$

Since the two terms are statistically independent, the covariance matrix is

$$R_{ij} \equiv <U_i U_j> = R_{ij}^{(v)} + R_{ij}^{(\varepsilon)} \ . \tag{5}$$

The velocity term can be written as

$$U_i^{(v)}(\vec{r}_o) = \int d^3 r \ W_{ij}(\vec{r}) \ v_j(\vec{r}_o + \vec{r}) \ , \tag{6}$$

where $W_{ij}$ is the tensor window function for the survey, given in fourier space by

$$W_{ij}(\vec{k}) = A_{im}^{-1} \sum_n \frac{\hat{r}_{n,m} \hat{r}_{n,j}}{\sigma_n^2 + \sigma_*^2} \ e^{i\vec{k}\cdot\vec{r}} \ . \tag{7}$$

Then the velocity part of the covariance matrix is the convolution of a window function and the velocity power spectrum

$$R_{ij}^{(v)} = \int \frac{d^3 k}{(2\pi)^3} \ \mathcal{W}_{ij}^2(\vec{k}) \ P_v(\vec{k}) \ , \tag{8}$$

where

$$\mathcal{W}_{ij}^2(\vec{k}) = W_{il}(\vec{k}) \ W_{jm}(\vec{k}) \ \hat{k}_l \ \hat{k}_m \ . \tag{9}$$

The velocity power spectrum is

$$P_v(k) \equiv <|v(\vec{k})|^2> = \frac{H^2 a^2}{k^2} P(k) \ , \tag{10}$$

where $P(k)$ is the density power spectrum. Like $A_{ij}$, $\mathcal{W}_{ij}^2$ and thus $R_{ij}$ are nearly diagonal.

As stated above, we consider power spectra from the IRAS–QDOT survey, the BBKS CDM ($\sigma_8 = 1$ $\Omega h = 0.5$) model, and HCDM simulations (normalized to COBE quadrupole $Q_2 = 17\mu K$). Since we anticipate our theoretical expectations to be low compared with the observed bulk flow, we will also include the IRAS–QDOT spectrum plus its $1\sigma$ error in our study. We have corrected the IRAS–QDOT and HCDM spectra for the redshift distortion pointed out first by Kaiser (1987).

The sparseness of the surveys we are considering will lead to features in the window function which will result in significant contributions to the bulk motion from scales much smaller than that of the survey. This is demonstrated in Fig. 1, where we show window functions for the LP and simulated catalogs in conjunction with the velocity power spectrum. Note that the window functions do not fall to zero outside of the central peak as they would if the volume were densely sampled. We also show a Gaussian $\exp(-k^2 R_*^2)$, where $R_*$ is the effective depth of the survey. For volume limited surveys of this type, the effective depth of the survey is $\approx R_{\max}/\sqrt{5}$, where $R_{\max}$ is the volume radius.

The noise term is

$$R_{ij}^{(\varepsilon)} = \left\langle \left( A_{il}^{-1} \sum_n \frac{\hat{r}_{n,l}\, \hat{r}_{n,i}}{\sigma_n^2 + \sigma_*^2} \right) \left( A_{jm}^{-1} \sum_{n'} \frac{\hat{r}_{n',m}\, \hat{r}_{n',j}}{\sigma_{n'}^2 + \sigma_*^2} \right) \right\rangle = A_{ij}^{-1} \;. \tag{11}$$

Thus the final covariance matrix [see eq. (5)] is

$$R_{ij} = \int \frac{d^3k}{(2\pi)^3}\, \mathcal{W}_{ij}^2(\vec{k})\, P_v(\vec{k}) + A_{ij}^{-1} \tag{12}$$

Given that $R_{ij}$ is nearly diagonal and isotropic, we can average over the three diagonal components and take

$$\sigma_1 \equiv \sqrt{R_{ii}/3} \;;\quad \sigma_1^{(v)} \equiv \sqrt{R_{ii}^{(v)}/3} \;, \tag{13}$$

to be the $1D$ variance for each of the three components of the bulk flow. Here a superscript $v$ refers to the error free quantities. The probability for measuring a bulk flow of magnitude

$V$ then becomes

$$P(V) = \left(\frac{2}{\pi}\right)^{1/2} \frac{V^2}{\sigma_1^3} \exp\left(-V^2/2\sigma_1^2\right) . \qquad (14)$$

The expectation value for the magnitude of the bulk flow in a given sample is

$$\Lambda \equiv \int_0^\infty V \, P(V) \, dV = \frac{4}{\sqrt{2\pi}} \sigma_1 \; ; \quad \Lambda^{(v)} \equiv \frac{4}{\sqrt{2\pi}} \sigma_1^{(v)}. \qquad (15)$$

Given $\sigma_1$, one can also use $P(V)$ to calculate the probability for observing a bulk flow greater than a given magnitude.

We construct three types of simulated cluster catalogs. Two with correlation length $r_o = 23$ and $14$ $h^{-1}$ Mpc, using a method similar to that of Postman *et al.* (1989) and Plionis *et al.* (1992), and a third where clusters are distributed randomly, *i.e.* $r_o = 0$. As usual we parametrize the uncertainty in our knowledge of the Hubble constant by $h = H/(100 \text{ km s}^{-1}\text{Mpc}^{-1})$. In order to mimic the LP catalog, we have eliminated all clusters with Galactic latitude $|b| < 15°$ and applied the csc extinction law $P(b) = dex\,[0.13(1 - \csc|b|)]$ (Strauss *et al.* 1993). For the 15,000 km/s samples, we chose 120 clusters as in the Lauer and Postman catalog. To construct catalogs with different radii we assume a constant density of clusters and scale the number of clusters accordingly.

## 3. Results

In Tables 1–3 we present our results for both the real and simulated cluster catalogs for the four power spectra we consider. For comparison with our results, we are interested in the uncorrected measurement of the bulk flow $V_{\text{LP}}$. This value is obtained by taking the magnitude of the bulk flow vector reported by Lauer and Postman in the cosmic rest frame, namely $V_{\text{LP}} = |\vec{F}| = |477 \text{ km/s}, -142 \text{ km/s}, 635 \text{ km/s}| = 807$ km/s. The bulk flow magnitude reported by LP, $689 \pm 178$ km/s, has been corrected for error bias. Note that this

'bias–free' magnitude is not the same as our 'error–free' magnitudes $\Lambda^{(v)}$ and they cannot be directly compared.

In Table 1 we present the results of our analysis applied to the LP catalog. We show both the raw ($\Lambda$) and the error–free ($\Lambda^{(v)}$) expectation values for the different power spectra. We also present the probability for measuring a bulk motion larger than that observed by LP, $P(V > V_{\rm LP})$. Results for the IRAS–QDOT spectra should be taken as lower bounds, due to a the lack of reliable information for $k < 0.02$ (see Feldman *et al.* 1993). These results indicate that the power spectra we have considered are inconsistent with the Lauer and Postman observation at the $90 - 94\%$ confidence level.

In Tables 2 and 3 we present the results for the simulated cluster catalogs. The results shown here are averaged over 50 realizations. The statistical variations in $\Lambda$ and $\Lambda^{(v)}$ were found to be of order 5% and so are neglected. In Table 2 we present the results for both $\Lambda$ and $\Lambda^{(v)}$ for the three classes of clustering properties and measurement errors (see Figure 2a–c). We see that the results of the highly clustered simulated catalogs are very similar to those from the LP catalog, suggesting that they have similar properties. Here we used $\beta = 16\%$, as reported by LP. The clustered catalogs enhance the bulk flow relative to random catalogs by $10 - 20\%$ (see Figure 2c).

Lauer and Postman, in collaboration with Strauss (Strauss 1993 private communication), are planning to undertake a project whereby they will survey all Abell clusters within $24,000$ km/s with improved measurement errors of $\beta = 10\%$. We have extended our analysis to include larger simulated surveys with the same cluster density. In Table 3 we show the results for a $25,000$ km/s survey for $\beta = 10\%$ (also see Figure 2a–c). These results indicate that the bulk flow in this extended survey is expected to be $\sim 250$ km/s.

For comparison, in Tables 2 and 3 and Figure 2d we show the results for the 'true' expected bulk velocity $\Lambda^{(t)}$ of the frame in the limit of infinite number of randomly placed sample objects for the HCDM and CDM power spectra.

## 4. Conclusions

We have presented a formalism to calculate the theoretical expectation for bulk flow in large scale surveys as a function of the geometry of the survey, its clustering properties and the assumed power spectrum. We apply the formalism to the LP survey as well as to simulated cluster catalogs. We find that there are two effects that significantly enhance the bulk motion for a catalog, namely sparseness and clustering. Sparse sampling of the velocity field leads to contributions from the velocity power spectrum on smaller scales than the survey. For a LP type survey this can result in a large enhancement over the expected motion of the volume as a whole. Clustering of the sample leads to even more enhancement, since it effectively increases the sparseness. For the clustering in the Abell catalog this results in an additional $10 - 20\%$ enhancement.

Our results indicate that the power spectra we have considered are inconsistent with the LP measurement of bulk flow at the $90 - 94\%$ confidence level. While this level of inconsistency is provocative, it does not indicate, as yet, a major problem with current theories of large scale structure formation. Improvements in the accuracy of the distance indicator should help clarify this issue.

We have also explored the convergence of the survey frame to that of the CMB frame as a function of radius using simulated cluster catalogs. Our results indicate that an Abell cluster survey of radius $25,000$ km/s with a distance measure accurate to $10\%$ will exhibit a bulk motion of order $250$ km/s, although this is somewhat uncertain due to a lack of information about the power spectrum on these scales. Extending the survey to this radius will provide interesting information about scales where the uncertainty in our knowledge of the power spectrum is large.

**Acknowlegements:** We are grateful to Nick Kaiser, Gus Evrard, and Andrew Jaffe for valuable comments. We also thank Tod Lauer and Mark Postman, and Michael Strauss for answering our questions regarding their work. HAF was supported in part by the National Science Foundation grant NSF–PHY–92–96020. RW was supported in part by NASA grant NAGW–2802.

– 10 –## REFERENCES

Bardeen, J. M., Bond, J. R., Kaiser, N. & Szalay, A. S. 1986, ApJ, 304, 15

Dressler, A., Faber, S. M., Burstein, D. Davies, R. L., Lynden–Bell, D., Terlevich, R. J. & Wagner, G. 1987, ApJ, 313, 42

Efstathiou, G. Dalton, G. B., Sutherland, W. J. & Maddox, S. J. 1992, MNRAS, 257, 125

Feldman, H. A., Kaiser, N. & Peacock J. A. 1993, ApJ in press

Feldman, H. A. & Watkins, R. 1993, proceedings of IAP Cosmic Velocity Fields, Ed. F. Bouchet & M. Lacheieze–Rey, in press

Kaiser, N. 1987, MNRAS, 227, 1

Kaiser, N. 1988, MNRAS, 231, 149

Klypin, A., Holtzman, J., Primack, J. & Regős, E. 1993, ApJ, 413, P48

Lauer, T. & Postman, N. 1993 preprint, LP

Plionis, M., Valdarnini, R. & Yi-Peng, J. 1992, ApJ, 398, 12

Postman, M., Spergel, D. N., Sutin, B., & Juszkiewicz, R. 1989, ApJ, 346, 588

Strauss, M., Cen, R. & Ostriker, J. P. 1993, proceedings of IAP Cosmic Velocity Fields, Ed. F. Bouchet & M. Lacheieze–Rey, in press
---

This preprint was prepared with the AAS LaTeX macros v3.0.

**Table 1: LP Catalog** ($V_{\rm LP} = 807$ km/s)

$\Lambda$ and $\Lambda^{(v)}$ are given in km/s.

| Spectrum | $\Lambda^{(v)}$ | $\Lambda$ | $\sigma_1$ | $P(V > V_{\rm LP})$ |
|---|---|---|---|---|
| IRAS | 205 | 470 | 295 | 0.06 |
| IRAS + $\sigma$ | 297 | 517 | 324 | 0.10 |
| CDM | 240 | 486 | 305 | 0.07 |
| MDM | 238 | 485 | 304 | 0.07 |

**Table 2:** $R_{\rm max} = 15000$ **km/s;** $\beta = 16\%$

$\Lambda$, $\Lambda^{(v)}$ and $\Lambda^{(t)}$ are given in km/s, $r_o$ values are in $h^{-1}$ Mpc.

|  | $r_o = 23$ | | $r_o = 14$ | | $r_o = 0$ | | |
|---|---|---|---|---|---|---|---|
| Spectrum | $\Lambda^{(v)}$ | $\Lambda$ | $\Lambda^{(v)}$ | $\Lambda$ | $\Lambda^{(v)}$ | $\Lambda$ | $\Lambda^{(t)}$ |
| IRAS | 193 | 479 | 172 | 448 | 156 | 442 | – |
| IRAS + $\sigma$ | 287 | 522 | 260 | 489 | 235 | 474 | – |
| CDM | 165 | 470 | 153 | 441 | 138 | 437 | 102 |
| MDM | 230 | 494 | 208 | 465 | 189 | 454 | 122 |

**Table 3:** $R_{\rm max} = 25000$ **km/s;** $\beta = 10\%$

$\Lambda$, $\Lambda^{(v)}$ and $\Lambda^{(t)}$ are given in km/s, $r_o$ values are in $h^{-1}$ Mpc.

|  | $r_o = 23$ | | $r_o = 14$ | | $r_o = 0$ | | |
|---|---|---|---|---|---|---|---|
| Spectrum | $\Lambda^{(v)}$ | $\Lambda$ | $\Lambda^{(v)}$ | $\Lambda$ | $\Lambda^{(v)}$ | $\Lambda$ | $\Lambda^{(t)}$ |
| IRAS | 100 | 228 | 89 | 218 | 78 | 213 | – |
| IRAS + $\sigma$ | 158 | 257 | 139 | 243 | 124 | 234 | – |
| CDM | 92 | 223 | 84 | 215 | 76 | 211 | 57 |
| MDM | 130 | 240 | 114 | 229 | 105 | 223 | 68 |

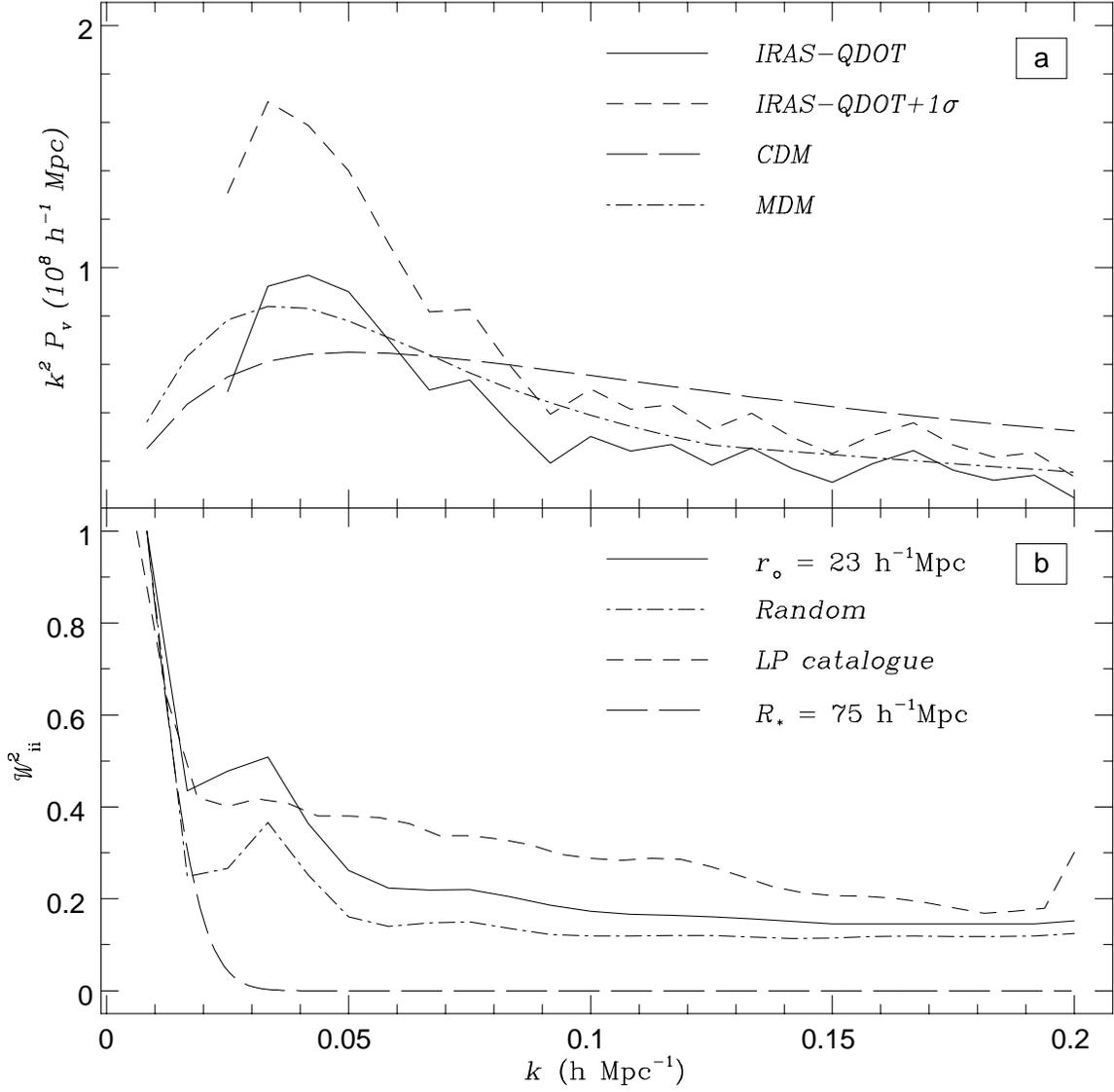

Figure 1

Fig. 1.— **Fig. 1)** In [a] we present the redshift corrected power spectra used in the analysis. In [b] we see the normalized tensor window for the highly clustered and random catalogs, the one for the LP catalogue and a guassian window function $\exp(-k^2 R_*^2)$.

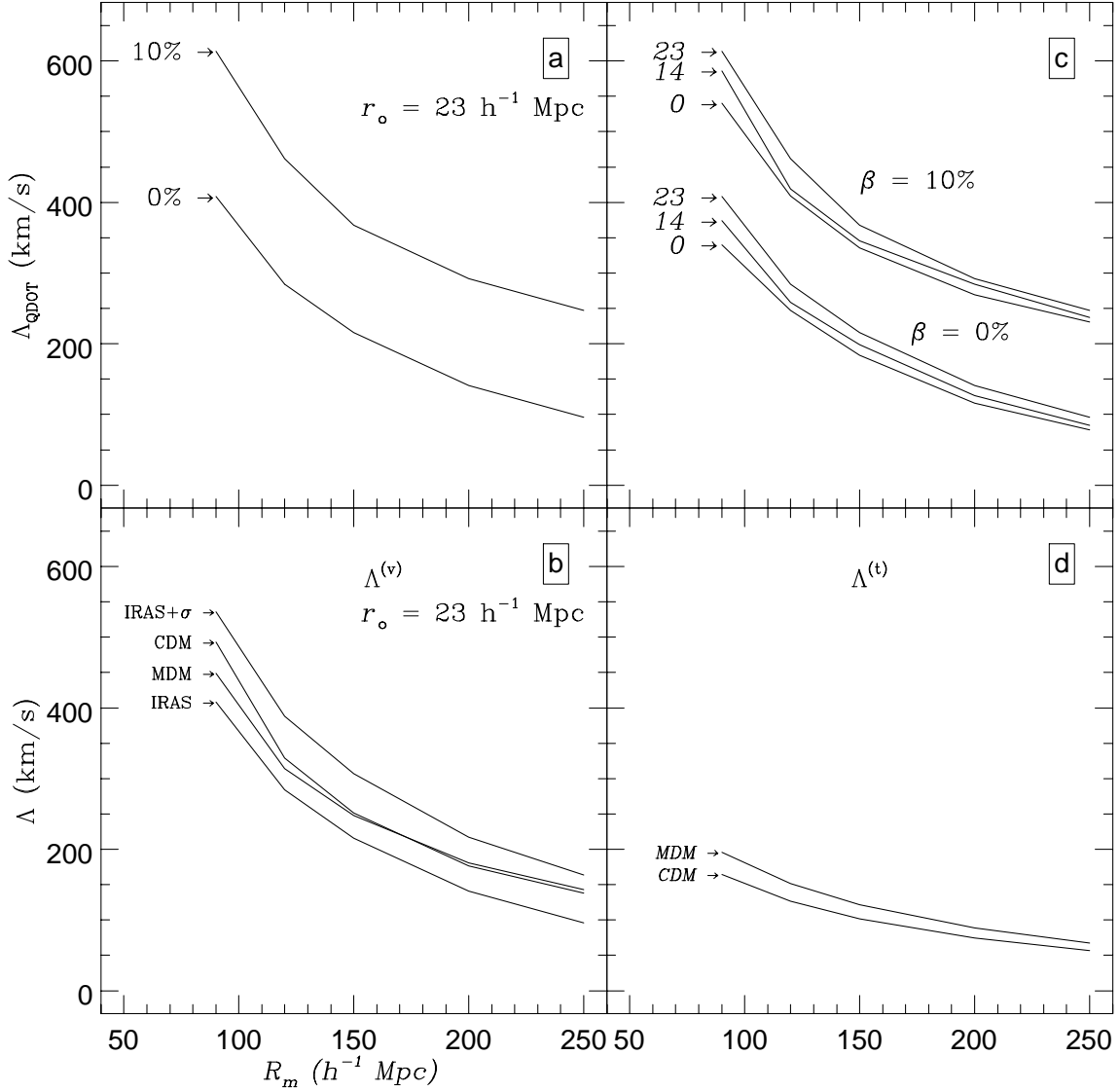

Fig. 2.— **Fig. 2)** In [a] we show the expectation values as a function of survey radius for $\beta = 10\%$ ($\Lambda$) and $0\%$ (error–free $\Lambda^{(v)}$) for the IRAS–QDOT power spectrum. In [b] we show the error–free expectation values ($\Lambda^{(v)}$) for the four power spectra we used. In [c] we show the expectation values for $\beta = 10\%$ ($\Lambda$) and the error–free ($\Lambda^{(v)}$) for random, moderately clustered and highly clustered (labeled $0, 14, 23$ respectively) simulations for the IRAS–QDOT power spectrum. In [d] we show the results for the limit of an infinite number of clusters for the CDM and HCDM power spectra ($\Lambda^{(t)}$).